# Seeing another Earth: Detecting and Characterizing Rocky Planets with Extremely Large Telescopes

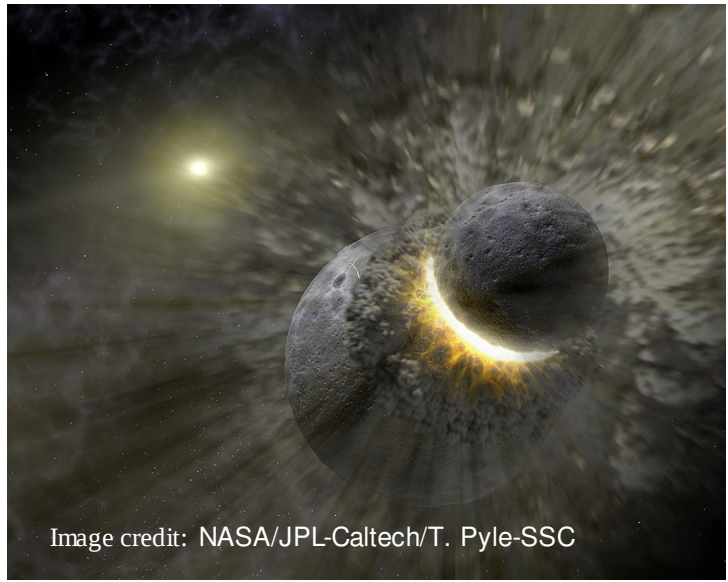

Image credit: NASA/JPL-Caltech/T. Pyle-SSC

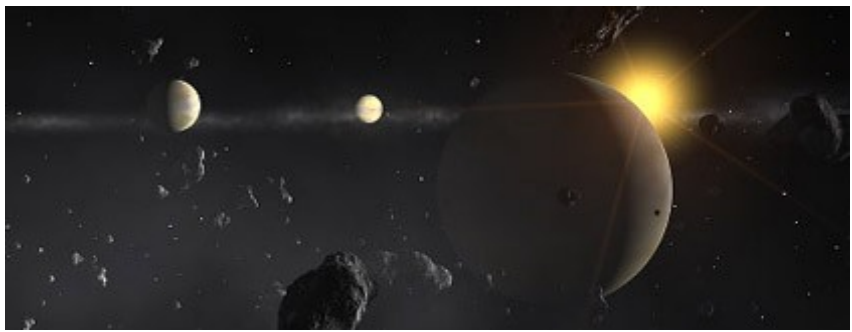

| | |
|---|---|
| Philip M. Hinz[1] | University of Arizona, |
| Scott Kenyon | Smithsonian Astrophysical Observatory, |
| Michael R. Meyer | University of Arizona, |
| Alan Boss | Carnegie Institute of Washington |
| Roger Angel | University of Arizona |

[1] Contact information: email: phinz@as.arizona.edu, telephone: (520) 621-7866

# 1. INTRODUCTION

In the past decade, astronomers have discovered nearly 1000 planetary systems. Microlensing, radial velocity, and transit surveys have yielded more than 300 planets with masses ranging from a few Earth masses ($M_E$) to 20-30 Jupiter masses ($M_J$). Observations in the thermal infrared with the IRAS and Spitzer satellites have revealed more than 500 debris disks with dusty material left over from the formation of Pluto-mass to Earth-mass planets. Although we have not discovered a planetary system similar to the Solar System, some stars have massive planets with orbits similar to those of Jupiter and Saturn. A few other stars show evidence for rocky planets with orbits similar to the terrestrial planets in the Solar System (Gould et al. 2006; Wyatt et al. 2007; Currie et al. 2008a).

These data demonstrate that planetary systems are common. Roughly 20% of solar-type stars have a Jupiter mass planet (Cumming et al. 2008). Another ∼ 30% may have super-Earths (Mayor et al. 2008b). More than half of the stars with masses of twice the Sun's mass have icy planets similar to Pluto and other large objects in our Kuiper belt (Currie et al. 2008b). Although current surveys are incomplete, at least 5% to 10% of solar-type stars have debris from the formation of Mars-mass or Earth-mass planets. These results suggest that we will eventually find a planetary system similar to our own. Today, exoplanet research is in the midst of a transition from discovery to analysis and understanding. Measurements of the mass-radius relation derived from transiting planets yield constraints on the internal structure and composition of gas giant planets (e.g., Guillot 2005). Data from HST and Spitzer now provide information on the amount of haze (Pont et al. 2008) and the day to night temperature gradient (Knutson et al. 2009) in the atmospheres of gas giants. These results suggest we are entering an age of 'comparative planetology,' where we will develop observational and theoretical toolkits to understand the frequency, formation, and evolution of planets and planetary systems.

The direct imaging of planetary systems is a potentially lucrative technique for carrying out detailed analysis of self-luminous planets. Initial discoveries in this area (Kalas et al. 2008; Marois et al. 2008) have already revealed surprising discrepancies between observational results and our current understanding of planet formation. For example, the three planets imaged around HR 8799 are surprisingly massive, and at large separations compared to what might be expected for a core-accretion model of planetary formation, suggesting formation from gravitational instability in a protoplanetary disk as a more likely possibility. The planet around Fomalhaut is best explained with a significant circumplanetary disk, an unexpected feature for a planet of several hundred Myr age.

Further direct detection with 8-10 m class telescopes and Extremely Large Telescopes (ELTs; apertures of >20 m assumed) will be important for detailed study of wide-period (>1 AU) gas giant planets. Direct imaging will allow determination of the radii, composition and internal energy of these planets. ELTs will likely be able to image some of the wider-period planets detected via radial velocity, providing important complementary information about these systems. This work, combined with observations of young (< 1 Gyr) gas giant planets at both NIR and thermal IR wavelengths, provides a strong science driver for ELT designs. Based on current models, we can expect to detect planets as faint as 0.5 $M_J$ around a sample of nearby stars. The detectable range of gas giant planets has been addressed elsewhere, and will not be discussed further in this white paper, although we note that such work is an important complement to the observations discussed below.

The limits set out for imaging self-luminous, gas-giant planets give the mistaken impression that anything smaller than Jupiter would be undetectable from the ground. However, it is worth noting that at 10 μm, where the Earth's emission peaks, it is nearly ten times brighter than Jupiter (Des Marais et al. 2002). **Planets that are emitting due to irradiation by their star may well be detectable even if they are significantly smaller than Jupiter.** Alternatively, rocky planets in formation may occasionally be heated by large collisions similar to the one thought to have generated the Earth-Moon system. The ideal wavelength for detection of such planets would likely be either L' or N band depending on the planet's temperature. A true Earth analog would favor detection in N band while a hotter Earth-like planet (~600 K) could be discovered through a search at L'. **The larger aperture ELTs being planned have the potential for providing direct detection in the IR of rocky planets similar to Earth.**

## 2. SCIENCE GOALS

Direct imaging of rocky planets provides important information about their size, temperature, composition, atmospheric structure, and ultimately, constrain the likelihood of their being habitable. We argue below that the detection of rocky planets with ELTs can be expected, based on our developing knowledge of the prevalence of lower mass planets and the expected performance of an ELT.

Simply detecting rocky planets with an ELT will be significant. Such observations will provide constraints on the range of conditions for possibly habitable planets. Specifically, we will want to understand the following:

- What are the sizes and temperatures of rocky planets?
- What can this information tell us about their composition and internal energy?
- What are their atmospheric compositions and structures?
- Are there any signs of greenhouse effects?

**By answering these questions, we can characterize the potential habitability of rocky planets and refining our understanding of this all-important criteria that allows us to begin addressing Astrobiology outside our solar system.**

The detection and detailed characterization of rocky planets lays the groundwork for carrying out comparative planetology. This will allow us to understand the formation of planetary systems and to learn how the internal structure and the surface properties of planets change as a function of their mass and age. Accomplishing these goals requires several basic quantities - the mass, radius, temperature, luminosity, and separation from the host star - of each planet. More detailed information - the spectrum - provides more constraints on the structure of the planet. Model atmospheres (to infer the internal structure and composition) and evolutionary calculations (to construct models for the formation and time evolution of the internal structure) link the observable quantities to the physical properties of the planet.

Achieving this broader goal requires a synergy between all of the observational tools used to discover and to characterize planetary systems. Microlensing, radial velocities, and transits provide the mass, radius, and orbital separation of planets. If the planet is luminous, transits can also yield the temperature and luminosity of the planet. In many cases, however, imaging is the only tool that provides the temperature and luminosity.

## 3. ELT EXOPLANET IMAGING

Successful imaging programs require high angular resolution and extraordinary sensitivity. Gas giant planets probably form at 3-20 AU. Current 6-m to 8-m class telescopes probe separations beyond 20-30 AU. Improving this limit by a factor of ten requires a larger aperture (e.g., ELTs of >20 m aperture) and better diffraction suppression (e.g., Kenworthy et al. 2007). At these separations, most planets are faint, more than a million - and usually more than a billion - times fainter than their parent star. Although the contrast is smallest at the longest wavelengths, angular resolution decreases with wavelength. Optimizing the contrast and resolution suggests imaging at 1.65 $\mu$m (H-band), where observations of reflected light from the planet have a reasonable chance of success, and at 3.8 $\mu$m (L'-band), where thermal emission from the planet is easiest to detect relative to the nearby star and large sky background (Hinz et al. 2006; Heinze et al. 2008).

These imaging programs have a complementary set of challenges. In the thermal infrared, minimizing the emissivity of the telescope and reaching the thermal background near the diffraction limit are the main technical goals. Current systems can achieve these goals at angular scales of 2-10 $\lambda$/D (Kenworthy et al. 2007; Apai et al. 2008). Thus, we choose a target resolution of 2 $\lambda$/D for background limited observations at L'.

In the near-infrared, achieving a stable point-spread-function (PSF) is more important than minimizing background emission. If the stability of the PSF allows long integration times (∼ 1 hr), then faint sources are detectable over some range of angular scales, n$\lambda$/D with $n \approx$ 2-10. For the example discussed below, we compare detection limits for a theoretical PSF using adaptive phase modulation (Angel et al. 2006) with adopted target values for the angular scale and contrast.

We envision several broad classes of ELT imaging programs that will yield vital information for improving our understanding of planets and planetary systems. For this discussion, we consider two specific programs that address the study of rocky planets possible with an ELT.

- ELT detections of young, molten proto-Earths will provide unique information on the origin of Earth and other terrestrial planets. We anticipate the discovery of 10 molten Earths around solar-type stars within 30-50 pc.

- ELT detection of super-Earth (3-30 $M_E$) mass planets around very nearby stars will provide important laboratories for understanding the limits of habitability on rocky planets. We anticipate being able to detect several rocky planets around stars within 8 pc.

**3.1. Molten Earths**

Throughout history, collisions have shaped the Earth and other planets in the Solar System. Mergers of several dozen lunar-sized to Mars-sized objects produced the proto-Earth 10-30 Myr after the formation of the Sun (Chambers 2001; Kenyon & Bromley 2006). Several Myr later, a giant collision with a Mars-sized planet removed the first terrestrial atmosphere and led to the formation of the Moon (Canup 2004b). During the Late Heavy Bombardment ∼ 3.8 Gyr ago, a sustained period of large collisions shaped the surfaces of all terrestrial objects (Koeberl 2003). Giant collisions were also probably responsible for the unusual composition of Mercury and the obliquity of Uranus.

Current theoretical estimates suggest that proto-Earths are luminous. Throughout the formation phase of 30-100 Myr, every large collision in the terrestrial zone produces a molten planet with a surface temperature of >1500 K and a fractional luminosity of ∼ $10^{-7} - 10^{-6}$ relative to its parent star. Although the cooling timescale for a molten planet in a vacuum is less than 0.1 Myr (Stern 1994), more detailed calculations suggest cooling times of up to a few Myr (e.g., Elkins-Tanton 2008). Because (i) the cooling time is a significant fraction of the formation time, (ii) each planetary system can have several Earth-mass planets, and (iii) each proto-Earth suffers several giant impacts, young proto-Earths are almost as easy to detect as the most luminous gas giants around the youngest stars!

Results from recent observational programs suggest that searches for proto-Earths are compelling (Mamajek & Meyer 2007). New results from microlensing (Bennett et al. 2008) and radial velocity (Mayor et al. 2008a,b) surveys suggest that super-Earths with masses of 5-20 $M_E$ are common. In addition to these direct detections, IRAS and Spitzer data show that 20% to 50% of young 1-2 $M_{sun}$ stars are surrounded by dusty disks of debris. This debris is almost certainly material left over from the process of planet formation. Thus, the statistics for the microlensing, radial velocity, and thermal infrared surveys suggest that many (if not most) stars have systems of planets with masses of at least 0.1-1 $M_E$.

To estimate the likelihood of detecting a molten proto-Earth around a nearby young solar-type star, we consider a simple model. We adopt an effective temperature of 1500 K and assume spectral energy distribution similar to a gas giant with the same temperature.

A molten proto-Earth then has absolute magnitudes $M_H \approx$ 17.3 at 1.6 $\mu$m and $M_L \approx$ 15.5 at 3.8 $\mu$m. A super-Earth with twice the radius of the proto-Earth is roughly 1.5 mag brighter. In this model, detecting molten proto-Earths around stars with d < 30 pc requires integration times of < 1 min at H and < 15 min at L for apertures of 20-30 m. Thus, an ELT could detect molten proto-Earths or super-Earths with separations exceeding 3 AU around stars with ages of 100 Myr at distances of 30 pc. We estimate there are ∼ 100 stars accessible to an ELT with ages ≤ 100- 200 Myr within 30 pc of the Sun. **If all 1-2 $M_{sun}$ stars form terrestrial planets like the Earth and proto-Earths remain molten and luminous for at least 10% of the formation epoch, an ELT survey of stars within 30 pc would yield ∼ 10 molten proto-Earths or super-Earths.**

Our currently poor understanding of young terrestrial atmospheres is the main uncertainty in this estimate. The detectability of a molten proto-Earth depends on the lifetime of the magma ocean at the surface, the evolution of the dense atmosphere, and the transfer of energy from the planet to the surrounding disk (e.g., Zahnle et al. 2007; Elkins-Tanton 2008; Miller-Ricci et al. 2009). Molten planets with no atmosphere are easy to detect but cool too quickly; planets with dense atmospheres remain hot for several Myr but are probably impossible to detect. Current 8-10 m telescopes may be able to detect hot planets that cool quickly but would need a large sample to have a reasonable probability of detection. **The importance of an ELT is that it would allow detection of planets that are fainter due to a thicker atmosphere. For these planets the surface stays molten for a much longer period, allowing a smaller sample of stars to detect such an object.** As predicted for young gas giants, our estimates assume a moderate atmosphere with windows that allow observations of the hot surface of the planet. Because several groups are working on this issue, our understanding is likely to improve considerably by the time an ELT is built.

## 3.2. Rocky Planets around Nearby Stars

As described above, a significant factor in favor of detecting rocky planets is their expected ubiquity compared to gas giant planets. Improved instrumental stability has allowed detection of planets with masses <10 $M_E$, using radial velocity (RV) measurements. These detections have occurred despite the strong selection effect against this due to the lower mass. Planets with minimum masses of 5-10 $M_E$ are now known around at least six stars (HD 40307b,c, Gl 581 c,d, Gl 876d, HD 181433b; from expoplanets.eu).. Similarly, detection of a 5 $M_E$ planet and a 13 $M_E$ planets through microlensing monitoring are a tantalizing indicator that massive planets may be common (Gould et al. 2006). Results from the Spitzer mission also support the notion that processes leading to terrestrial planet formation are common around sun-like stars (e.g. Siegler et al. 2006, Meyer et al. 2008). Certainly the current data is not sufficient to say, definitively, that rocky planets are common, but continued precision improvements in RV methods and additional events for microlensing, as well as results from the Kepler mission, could confirm that rocky planets are significantly more common than giant planets. If these preliminary indicators of high frequency are correct, the detection of a terrestrial planet will be feasible with ground-based ELTs.

The detection of rocky planets with an ELT is likely to be easiest at either L' (3.8 $\mu$m) or N (10.5 $\mu$m) band, depending on the temperature of the planet. These passbands provide relatively good sensitivity, due to the high transparency of the atmosphere at these wavelengths.

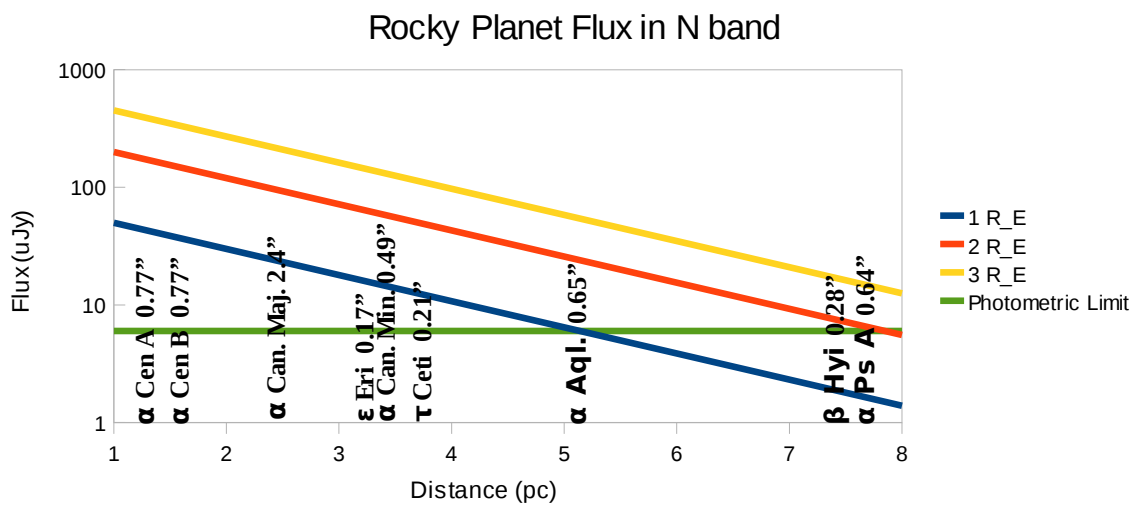

**Figure 1.** N band brightness of a T=280K planet versus parent star distance for planets of size 1 (blue), 2 (red), and 3(yellow) Earth radii (R_E). Candidate stars for a southern hemisphere site are listed at their respective distances. The habitable zone (T=280 K) angular separation for each star is listed. Only those stars with habitable zones > 2 $\lambda$/D for an ELT of 25 m are listed. A photometric limit of 6 $\mu$Jy is assumed.

Observations at N band would need to achieve a contrast of $10^{-7}$ and be capable of detecting a planet with an absolute magnitude $M_N \approx$ 20.0 (0.5 µJy; Tinetti et al. 2006 ). Super-Earths might plausibly be up to a factor of ten brighter than this estimate( $M_N \approx$ 17.5, or 5 µJy). Based on current sensitivity limits from 6-8 m telescopes, an IR-optimized ELT (one with low telescope emissivity) could plausibly reach a sensitivity of 20 µJy in an hour, or 6 µJy in a 10 hour observation.

The expected angular separation of a rocky planet with a peak flux near 10 µm wavelength is simply a function of the apparent magnitude of the star. Thus, the most attractive stars to search are also the brightest ones in the sky. **If we follow the above assumption that rocky planets are a common occurrence(P>30%), then surveying the ten best candidates with an ELT provides a high probability of detecting a rocky planet.** Figure 1 estimates the brightness of rocky planets of size 1-3 $R_E$ around stars from 1-8 pc. Candidate stars are shown on the plot with the expected separation of a planet in the star's habitable zone. The planet flux for N band was taken from models in Des Marais et al. (2002). The expected angular separation of a planet with a temperature of 280 K is calculated by taking into account the luminosity and distance of the star. Planets around all the listed stars that correspond to a habitable zone > 2 $\lambda$/D for a 20-30 m ELT observing at N band.

Of course, rocky planets may be at a range of separations about their parent stars. Planets further out, and thus cooler, would likely be undetectable. However, planets closer to their star, with equilibrium temperatures correspondingly higher, could very well be bright enough to study. An Earth-size planet at an equilibrium temperature of T=600 K would be roughly four times closer than a true Earth analog (at 280 K) and would have a peak flux in the L band atmospheric window. If we assume blackbody emission of the planet at its equilibrium temperature it would have an absolute magnitude of $M_L \approx 20.0$ (2 µJy). This would result in a contrast of $10^{-7}$ with its parent star. Figure 2 shows the expected flux of hot rocky planets of various sizes and for a range distances of the parent star. These planets would have an angular separation much closer to the star. Candidate stars for observation are listed in the Figure at their respective distances. A photometric limit of 0.5 µJy, expected for an ELT limited by sky and telescope background in a one hour observation is also shown.

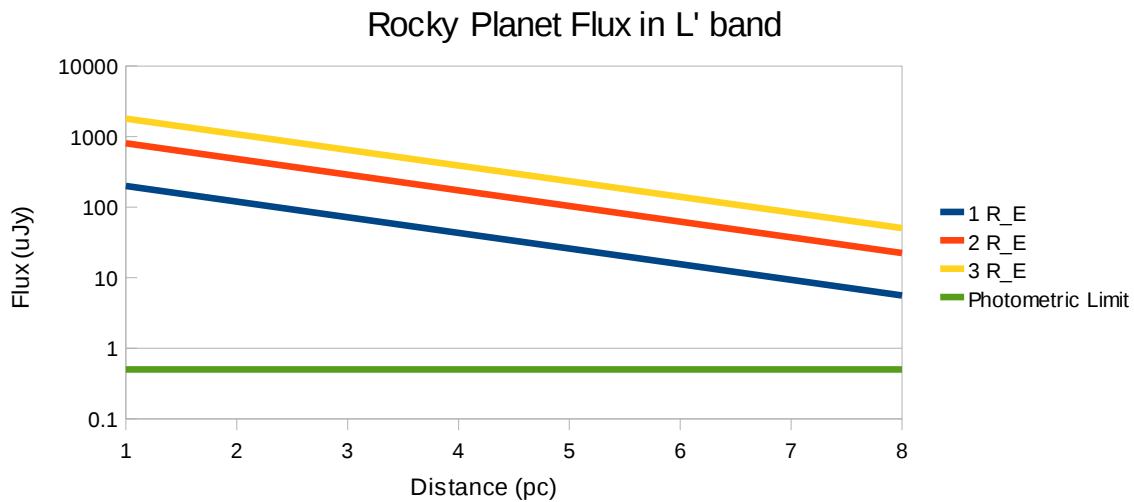

**Figure 2.** L' band planet brightness of a T= 600 K planet versus parent star distance for planets of size 1 (blue), 2 (red), and 3(yellow) Earth radii (R_E). Candidate stars for a southern hemisphere site are listed at their respective distances. The angular separation for a T=600 K planet for each star is listed. Only those stars with T=600 K zones > 2 λ/D for an ELT of 25 m are listed. A photometric limit of 0.5 µJy is assumed.

The approximation of a 600 K blackbody for a hot Earth is almost certainly inaccurate. However, SED models of hot planets with a range of atmospheric compositions have been studied by Miller-Ricci et al. (2009). The models suggest that the L' band is a generally transparent window, providing support that detection of hot Earths may be feasible with ELTs.

It is worth noting that the limiting factor for the sample size of "warm" versus "hot" Earths is different. For the N band detections, photometric sensitivity limits the number of stars where a rocky planet would be detectable. For "hot" Earths, the photometric sensitivity will, in principle, allow detection of even sub-Earth mass planets. However, the angular separation of such objects is only outside of ~2 λ/D of an ELT for the sample of objects listed in Figure 2. **For detection of an exo-Earth at either L' or N band, the sample size is modest, but the potential for discovery and detailed follow-up is dramatic.**

The technical challenges to accomplish such observations are significant, but tractable. Primarily, a large aperture that is optimized for infrared observations is important, suggesting a need for an adaptive secondary mirror or similar way of preserving a low background telescope design. Both the L' and N band atmospheric windows are quite good even from lower elevation observatories, but a high, dry site would be ideal for these observations. Finally an optimized diffraction suppression approach and precision AO system will be critical to achieve the needed contrasts at an angular distance of several λ/D. Development and on-sky testing of these techniques will need to be an important component of extrasolar planet technology development to take advantage of the large light grasp of future ELTs.

The above observations argue the need for a thermal infrared capability on an ELT. Before an ELT begins science operations, it is likely that JWST will have been operating for several years. No ground-based facility will compete with JWST for raw sensitivity from 0.8-25 microns. However, an ELT will have more than x4 finer spatial resolution and a superior PSF for high contrast imaging (due to mid-frequency surface errors and subtle alignment issues for JWST). This capability will be particularly enabling for direct imaging of rocky planets in or close to the habitable zone.

## 4. SUMMARY

In the last two decades, the number of candidate and confirmed exoplanets has grown from a few to over 300. In the next decade, microlensing, radial velocity, and transit techniques will probably increase this number by another factor of 10-100. Thus, by the time an ELT has first light, exoplanet science will be dominated by comparative planetology, where our goal will be to use observational and theoretical tools to understand the internal structure of planets and to place better limits on the mechanisms for the formation and evolution of planetary systems. **Direct imaging of rocky planets has the potential for providing some of the most detailed information about these objects which will be crucial in understanding their physical makeup, their formation, and, ultimately, their habitability.** ELTs, properly designed, instrumented, and operated, can play an enabling role in these discoveries.